# Doppler Fluctuation Spectroscopy of Intracellular Dynamics in Living Tissue

ZHE LI,[1,*] HAO SUN,[1] JOHN TUREK,[2] SHADIA JALAL,[3] MICHAEL CHILDRESS,[4] AND DAVID D. NOLTE[1]

[1]*Department of Physics and Astronomy, Purdue University, 525 Northwestern Ave, West Lafayette, IN 47907, USA*
[2]*Department of Basic Medical Sciences, Purdue University, 625 Harrison Street, West Lafayette, IN 47907, USA*
[3]*Department of Medicine, IU School of Medicine, 535 Barnhill Drive, Indianapolis, IN 46202, USA*
[4]*Department of Veterinary Clinical Sciences, Purdue University, 625 Harrison Street, West Lafayette, IN 47907, USA*
*\*zheli@purdue.edu*

**Abstract:** Intracellular dynamics in living tissue are dominated by active transport driven by bioenergetic processes far from thermal equilibrium. Intracellular constituents typically execute persistent walks. In the limit of long mean-free paths, the persistent walks are ballistic, exhibiting a "Doppler edge" in light scattering fluctuation spectra. At shorter transport lengths, the fluctuations are described by lifetime-broadened Doppler spectra. Dynamic light scattering from transport in the ballistic, diffusive or the cross-over regime is derived analytically, including the derivation of autocorrelation functions through a driven damped harmonic oscillator analog for light scattering from persistent walks. The theory is validated through Monte Carlo simulations. Experimental evidence for the Doppler edge in 3D living tissue is obtained using biodynamic imaging based on low-coherence interferometry and digital holography.

## 1. Introduction

Motions inside living cells, and motion of the cells themselves, are ubiquitous signatures of the active processes involved in the maintenance of cellular function and health. Many aspects of cellular function involve active movement driven by energetic processes. Conversely, thermal motions, though participating in subcellular processes such as molecular diffusion and membrane flicker, are physical processes that continue after the death of the cell. Therefore, driven motion is a defining characteristic of living matter. Quantifying the many aspects of active cellular motions provides a measure of cellular health or a measure of deviation from normal behavior caused by disease or by applied xenobiotics. Cellular dynamics become surrogates that can be used as real-time endogenous reporters in place of nonendogenous fluorophores or end-point measurements [1, 2] when studying how tissues respond to changing environments or to applied therapies.

Motions in two-dimensional cell culture are easily observed under a microscope as physical displacements. Furthermore, adaptive optics combined with light sheet microscopy and lattice light sheet microscopy can achieve 3D *in vivo* aberration-free imaging of subcellular processes [3, 4]. However, dynamic light scattering (DLS) and Doppler fluctuation spectroscopy are better suited for ensemble measurements of a broad range of intracellular motions across a wide field of view. These ensemble techniques are sensitive to motion changes and can be used to monitor cellular health, disease progression and drug response. Spatial localization in DLS in tissue can be achieved with low coherence [5-7], including dynamic signals observed in optical coherence tomography (OCT) [8] and optical coherence imaging (OCI) [9] which is a full-frame form of OCT [10, 11]. Diffusing-wave spectroscopy (DWS) [12] and diffuse correlation spectroscopy (DCS) are techniques that apply DLS to media like tissues. DCS is widely used for cerebral blood flow monitoring [13], in which temporal autocorrelation functions of speckle or light electric fields are measured, analyzed, and then compared with a certain type of motion, e.g. Brownian motion. Autocorrelation functions for molecular Brownian motion are usually measured in the time range of $10^{-5}$ to $10^{-2}$ seconds [14], which covers a significantly different range than active transport processes. As another example, diffuse reflectance spectroscopy (DRS) uses a spectrometer to analyze the diffuse reflectance of a tissue illuminated by a broadband light source and quantifies optical and physiological properties of tissues [15, 16].

Biodynamic imaging [17-19] is an optical imaging technology derived from OCI, with enhanced partially-coherent speckle generated by broad-area illumination with coherence detection through digital holography [20-23]. Biodynamic imaging penetrates up to 1 mm into living tissue and returns information in the form of dynamic light scattering across a broad spectral range [6, 24]. Frequency-domain decomposition of the light fluctuations using tissue dynamics spectroscopy (TDS) [18, 25] produces broad-band fluctuation spectra that encompass the wide variety of subcellular motions. When pharmaceutical compounds are applied to a tissue, dynamic cellular processes are modified, and these modifications appear as changes in the fluctuation spectra, which can help provide information about the effect of the compound on cellular processes such as necrosis and apoptosis [25]. This type of phenotypic profiling has seen a resurgence in recent years as a more systems-based approach to drug discovery and development [26].

This paper focuses on lifetime-broadened Doppler scattering from persistent walks. We present evidence that shows Doppler fluctuation spectra from midsections of 3D cultured tissues as the sum of active intracellular processes with long persistence distances, i.e. in the ballistic regime, which is consistent with findings from motion tracking within 2D tissues. Our model is based on random walks with a simple exponential distribution of free path lengths, where a particle walks at a constant velocity along a mean-free path. This "piecewise continuous random walk" model leads to a temporal cross-over from ballistic transport at short time scales to diffusive transport at long time scales. The approach is fully statistical, without resolving individual scattering objects, by restricting the analysis to ensembles of actively transporting subcellular constituents. The theory of light scattering from random walks is developed for field-based heterodyne detection. Transport in the ballistic, diffusive and the cross-over regime is described analytically, including the derivation of autocorrelation functions in the two limits and a driven damped harmonic oscillator model for persistent walks in the cross-over regime. The theory is validated by Monte Carlo simulations. Experimental measurements of Doppler fluctuation spectra, obtained using tissue dynamics spectroscopy on living tissue culture and living cancer biopsies are presented, followed by a general discussion on the potential applicability of Doppler fluctuation spectroscopy for drug screening.

## 2. Materials and Methods

### 2.1 Persistent Walk

Many biological applications proceed via active persistent walks that have persistent motions of relatively uniform speed $v_0$ travelling a mean-free length (also known as the persistence length) $L_p$ in a mean-free time $t_p$ (also known as the persistence time) before changing direction or speed. Persistent walks have two opposite limiting behaviors. When the persistence time is much longer than an observation time, then the transport can be viewed as an ensemble of ballistically transported objects. This is the ballistic limit. When the persistence time is much shorter than an observation time, then the transport approaches a Wiener process. The Wiener process has a path that is nowhere differentiable [27]. This is the diffusion limit, although, in the case of active media, it is active diffusion that significantly exceeds thermal Brownian motion. The ballistic limit and the diffusion limit have well-recognized properties in terms of dynamic light scattering. However, many biological transport processes occur in the cross-over regime between these extremes.

The key parameters characterizing the walks are the mean-squared speed during the free runs and the mean-free time between changes in speed or direction. A model that describes this process of free runs with mean persistence times is called the Ornstein-Uhlenbeck process [27] given by

$$dv = -\gamma v dt + \Gamma dW_t \qquad (1)$$

for one-dimensional transport, where $1/\gamma = t_p$ is the persistence time, $\Gamma$ is the amplitude of the fluctuations, and $dW_t$ is a Wiener process of unit variance. Setting $x(0) = 0$, the associated position process is described by

$$x(t) = \frac{v_0}{\gamma}\left[1 - \exp(-\gamma t)\right] + \Gamma \int_0^t dt' \exp(-\gamma t') \int_0^{t'} dW_{t''} \exp(\gamma t'') \qquad (2)$$

The mean-squared displacement (MSD) for quasi-ballistic transport in 1D is

$$\langle x^2(t) \rangle = \frac{\Gamma^2}{\gamma^2} t + \frac{v_0^2}{\gamma^2}\left[1 - \exp(-\gamma t)\right]^2 - \frac{\Gamma^2}{2\gamma^3}\left[3 - 4\exp(-\gamma t) + \exp(-2\gamma t)\right] \qquad (3)$$

where $v_0$ is the molecular motor speed. In the long-time limit, this is

$$\langle x^2(t) \rangle = \frac{\Gamma^2}{\gamma^2} t = 2v_0^2 t_p t = 2Dt \qquad (4)$$

where the effective diffusion coefficient is $D = v_0^2 t_p$ related to the speed and the persistence time, but unrelated to temperature or thermal processes. The relationship in Eq. (4) establishes the fluctuation-dissipation theorem for active transport

$$\Gamma^2 t_p = 2v_0^2 \qquad (5)$$

that relates the persistence time and speed to the magnitude of the fluctuations. Based on this relation, the MSD in Eq. (3) is expressed in terms of the mean-free path length $L_p = v_0 t_p$ as

$$\langle x^2(t) \rangle = 2L_p^2 \frac{t}{t_p} - 2L_p^2\left[1 - \exp\left(-\frac{t}{t_p}\right)\right] \qquad (6)$$

The MSD is plotted in Fig. 1 for several values of mean-free path. At short times, the MSD grows as the square of time, which is representative of ballistic transport, while for time $t > 2t_p$ the MSD grows linearly in time, which is representative of diffusive transport. Therefore, the MSD displays a temporal transition from ballistic to diffusive transport depending on the observation time.

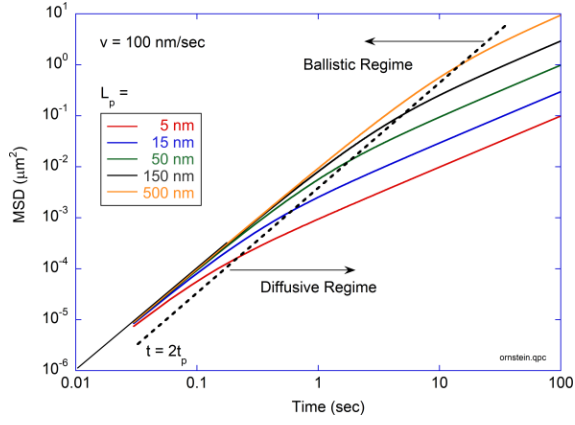

Fig. 1 Average mean-squared displacement as a function of time for an Ornstein-Uhlenbeck process for a family of $L_p$ values with a fixed speed of 100 nanometers per second. The transition from the ballistic to the diffusion regime occurs for $t = 2t_p$ along the dashed line.

## 2.2 Doppler Light Scattering

The light-scattering configuration for dynamic light scattering from a moving particle is shown in Fig. 2. The incident light has an initial $k$-vector $\mathbf{k}_1$ that is scattered by a small particle into a final $k$-vector $\mathbf{k}_2$. The momentum transfer in the scattering process is $\mathbf{q} = \mathbf{k}_2 - \mathbf{k}_1$, where the magnitude of the transferred momentum is

$$|\mathbf{q}| = k\sqrt{2(1-\cos\theta)} = 2k\sin(\theta/2) \tag{7}$$

at the scattering angle $\theta$. The Doppler frequency shift from the central frequency of the incident photon is given by

$$\Delta\omega_\phi = \mathbf{q} \cdot \mathbf{v} = qv\cos\phi = \omega_D \cos\phi \tag{8}$$

where

$$\omega_D = qv \tag{9}$$

is the maximum (or co-linear) Doppler angular frequency shift, $v$ is the velocity of the particle, and $\phi$ is the angle between the particle velocity and the momentum transfer vector. For forward scattering, $\theta = 0$, and the Doppler frequency shift is identically zero. For backward scattering, the momentum transfer $q = 4\pi n / \lambda_0$ is a maximum, and the Doppler frequency shift depends only on the particle velocity through $\omega_D = qv$.

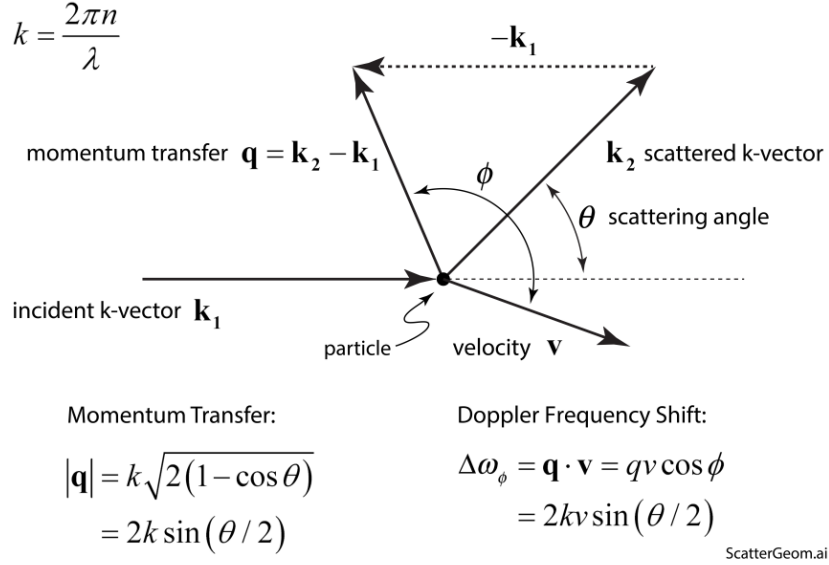

Fig. 2 Doppler scattering geometry for incident and scattered *k*-vector, *q*-vector and particle velocity. The scattering angle of the light is $\theta$, and the angle between the *q*-vector and the particle velocity is $\phi$.

The co-linear Doppler frequency shift $\omega_D$ and the persistence time $t_p$ (and equivalently the momentum transfer $q$ and the mean free path $L_p$) set a dimensionless scaling parameter that divides the ballistic transport regime from the diffusive regime. The dimensionless parameter is called the Doppler number, or the $N_D$, given by

$$N_D = \omega_D t_p = qL_p \qquad (10)$$

The characteristic scale is set when $N_D = 1$

$$N_D = \frac{L_p}{\lambda_{\text{red}}} = 1 \qquad (11)$$

which is defined in terms of the reduced wavelength as

$$\lambda_{\text{red}} = \frac{\lambda_0}{4\pi n} \approx 50 \text{ nm} \qquad (12)$$

for a refractive index $n \approx 1.35$ and a free-space wavelength $\lambda_0 = 840$ nm. Therefore, the dividing line between diffusive transport and ballistic transport occurs when the mean-free path is greater than approximately 50 nm. The conditions on the $N_D$ for the different regimes are

$$N_D > 3 \quad \text{Doppler Regime}$$
$$N_D < \frac{1}{3} \quad \text{Diffusion Regime}$$

(13)

although the division is not sharp. Doppler effects dominate when $N_D > 3$, and diffusion effects dominate when $N_D < 0.3$. Most active transport processes in cells have mean-free paths larger than this, placing most active subcellular processes in the Doppler regime (see next section).

*2.3 Processive Motion in Biological Processes*

Active intracellular transport is processive, meaning that motion persists for multiple cycles of ATP or GTP hydrolysis [28]. For molecular motors, the step length is fixed at $\delta_{ATP}$ per hydrolysis (e.g. for kinesin $\delta_{ATP} = 8$ nm [29]), with a mean value of $n$ steps before the motor detaches. The mean-free path for the persistent motion is then $\Delta = n\delta_{ATP}$. Likewise, for cytoskeletal restructuring, periods of protrusion are interspersed with periods of retraction, with characteristic mean-free lengths. Examples of intracellular dynamics, speeds and lengths are given in Table 1 for a variety of motions under a variety of conditions [29-40]. For these processes, the Doppler frequency depends on the observation wavelength and observation direction. The Doppler frequencies in Table 1 are calculated for a backscattering configuration using a free-space wavelength of $\lambda_0 = 840$ nm. Speeds range from several microns per second (organelles or vesicles carried by molecular motors) to several nanometers per second (cell membranes driven by cytoskeletal processes). The corresponding Doppler frequencies (maximum frequencies in a backscatter configuration) are tens of Hz to tens of mHz. The mean Doppler frequency (averaged over many cellular volumes in living tissue) is zero because transport is isotropically averaged over all directions. For these processive processes associated with kinesin, dynein, myosin V, cytoskeleton restructuring, and filopodia and lamellipodia, the Doppler numbers in Table 1 are greater than unity and can range into the hundreds. Therefore, processive motors and cytoskeletal restructuring are in the Doppler regime. An interesting case is for kinesin/dynein complexes, which are engaged in a tug-of-war transporting vesicles in alternating directions on the microtubules. The $N_D$ is smallest for this case in Table 1, and is in the cross-over regime.

**Table 1 Speeds, Doppler Frequencies and Mean Free Path Lengths**

| Motor or polymerization | Speed | Doppler Frequency | Distance or Time | $\omega_D t_p$ or $qL_p$ | Ref |
|---|---|---|---|---|---|
| Kinesin | 2 μm/sec | 6 Hz | | | [30] |
| Kinesin | 1 μm/s | 3 Hz | | 15 | [29] |
| Kinesin | 800 nm/s | 2.7 Hz | | | [31, 32] |
| Kinesin | 1 μm/s | 3 Hz | 10 sec | 200 | [33] |
| Kinesin | | | 1 micron | 20 | [34] |
| Kinesin | 1 μm/s | 3 Hz | 600 nm | 10 | [35] |
| Kinesin/Dynein | 800 nm/s | 2.7 Hz | 100 nm - 300 nm | 2 - 6 | [36] |
| Dynein/Dynactin | 700 nm/sec | 2.1 Hz | 1 | 20 | [41] |
| Myosin V | 300 nm/s | 1 Hz | 1.6 microns | 30 | [37] |
| ParA/ParB | 100 nm/s | 0.3 Hz | 2 microns | 40 | [38] |
| Actin network polymerization | 5 nm/s | 0.02 Hz | | | [39] |
| Tubulin polymerization | 20 nm/s – 300 nm/s | 0.07 – 1 Hz | 300 sec | 15-100 | [40] |
| Filopodia extending | 40 nm/s | 0.12 Hz | 130 sec | 100 | [42] |
| Filopodia retracting | 10 nm/s | 0.03 Hz | 100 sec | 20 | [42] |

## 2.4 Dynamic Spectroscopy of Living Tissue

Fluctuation frequencies relate to Doppler frequency shifts caused by light scattering from the subcellular constituents that are in motion, creating beats among all the multiple partial waves. The speeds of intracellular dynamics range across three orders of magnitude from tens of nanometers per second (cell membrane) [43-46] to tens of microns per second (organelles, vesicles) [47-50]. For near-infrared backscattering geometry, these speeds correspond to Doppler frequencies from 0.01 Hz to 10 Hz. Because of the wide variety of intracellular processes and the wide range of speeds, the fluctuation spectra obtained from dynamic light scattering of living tissue contain a continuous distribution of Doppler-broadened spectra. The experimentally-measured field-based fluctuation spectrum is

$$S'_E(\omega) = \int_0^\infty \rho(\omega_D) S_E(\omega, \omega_D) d\omega_D \qquad (14)$$

where $S_E(\omega, \omega_D)$ is the fluctuation spectrum of each individual process and $\rho(\omega_D)$ is a normalized distribution function that captures the range of intracellular Doppler processes. The combined power spectral density of Eq. (14) produces an envelope that contains the individual Doppler spectra of the underlying processes. For this reason, the power spectral density from most living tissue samples has a broad frequency dependence without a distinct Doppler edge. However, it is sometimes possible to observe from experiments either sharpening or broadening of an underlying Doppler edge.

## 2.5 Experimental Setup

Spectroscopic responses of several types of biological samples were measured and analyzed using the 6-F biodynamic imaging system shown in Fig. 3. A Superlum S840-B-I-20 superluminescent diode (SLD), with a center wavelength at 840 nm and full power output of 15 mW, was used as the light source. The SLD has a short coherence length of approximately 10 microns, enabling the formation of low-coherence holograms in a Mach-Zehnder interferometric configuration with a CCD camera as the detector at the Fourier plane. A Q-Imaging EMC2 camera captures 500 frames at 25 fps and 50 frames at 0.5 fps, and a stitching algorithm is used to construct a continuous spectrum ranging from 0.01 Hz to 12.5 Hz [51]. Holograms [Fig. 4(a) with a close-up in Fig. 4(b)] are written by scattered photons that share the same optical path length (OPL) as the reference arm. By adjusting the delay stage, light scattered from different depths inside the sample can be selected, setting the "coherence gate" for the detection. Tissue samples are typically between 0.5 mm and 1 mm thick, and the coherence gate is typically set at about 200 to 500 microns inside the sample. The transport length of light in many types of tissue samples is approximately 100 microns. Therefore, the light selected by the coherence gate in our experiments is multiply scattered with between 4 to 10 high-angle scattering events. Multiple scattering compounds the Doppler shifts and broadens the fluctuation spectra. The digital holograms are reconstructed numerically using a 2D FFT to generate optical sections approximately 400 microns inside the tissue. A reconstructed image and its conjugate are shown in Fig. 4(c) with a close-up in Fig. 4(d).

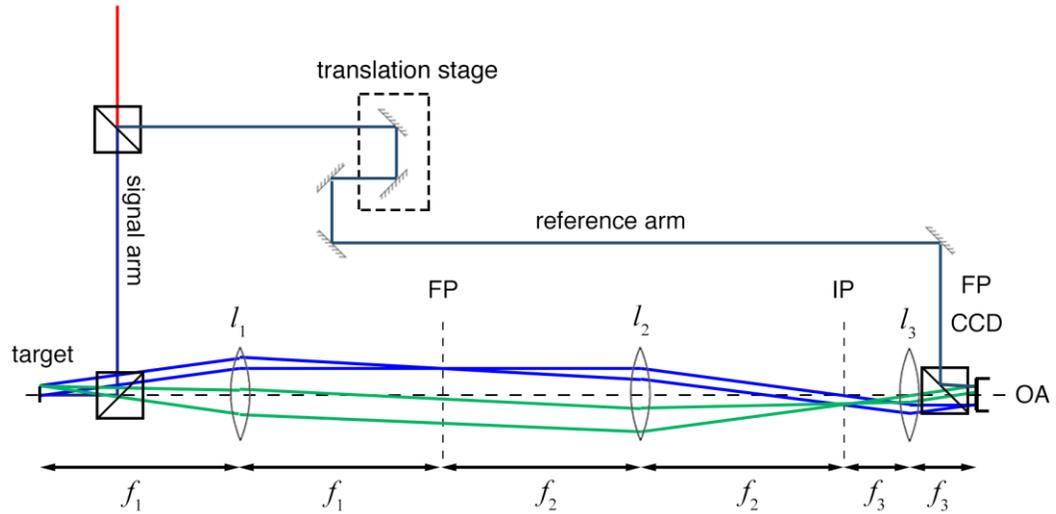

Fig. 3 Biodynamic imaging system in a Mach-Zehnder configuration. The camera is located on the Fourier domain of the sample. A translation stage is used to select the coherence gate and to form images at different depths inside the sample.

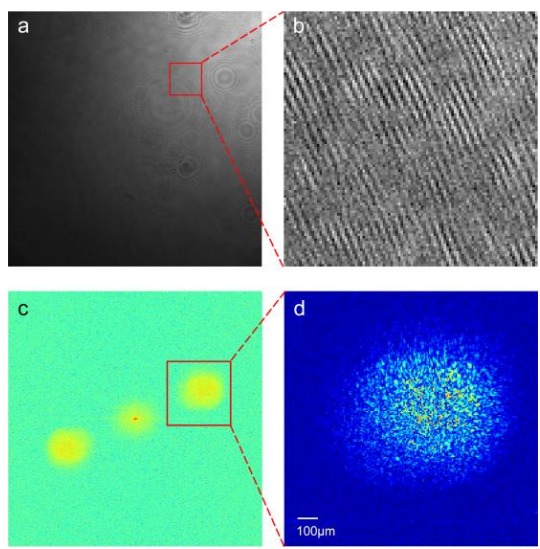

Fig. 4 An example of reconstruction of holograms captured by the BDI system. (a) Raw hologram image, (b) fringes, (c) FFT of the hologram and (d) sample image in the first order

*2.6 Sample Preparation*

Two types of tumor tissue were examined experimentally and analyzed for Doppler features in this paper: tumor spheroids and tumor biopsies. Multicellular tumor spheroids (MCTS) are small clusters of cancer cells grown *in vitro*. The three-dimensional growth of the spheroids captures many of the microenvironmental features of naturally occurring tumor tissues, including extracellular matrix and cell-to-cell contacts [52]. Tumor biopsies are even more biologically and physiologically relevant than 3D tissue culture and are obtained from living patients (animal or human with approved IRB) either through surgical resection or by needle core biopsies.

For the tumor spheroids, cell lines were obtained from American Type Culture Collection (ATCC), Manassas, Virginia, and cultured at 37°C in a humidified $CO_2$ incubator. HT-29 cells were cultured in McCoy's 5a Medium and the MIA-PACA2 were cultured in Dulbecco's Modified Eagle's Medium. All media contained 10% fetal calf serum (Atlanta Biologicals), penicillin (100 IU), and streptomycin (100 μg/mL). Tumor spheroids were created by seeding a 50 mL rotating bioreactor and growing the cells for 7-14 days until 400-600 μM diameter spheroids were formed. Spheroids were immobilized in a thin layer of 1% low gel temperature agarose (Sigma-Aldrich Chemical Co) made up with the basal medium in 96 well tissue culture plates. DLD-1 samples were grown as 3D tumor spheroids using Corning U-bottom spheroid plates. Cells are incubated in a 96-well plate and immobilized with low gel-temperature agarose. Esophageal tumor biopsies were collected and transported in chilled RPMI-1640 medium supplemented with HEPES (Gibco). Within 2 hours of collection, small pieces 1 mm or less were cut from the biopsy and immobilized in low gel temperature agarose in basal medium similar to the tumor spheroids. Canine B-cell lymphoma biopsies were handled with the same procedures [53].

## 3. Results

### 3.1 Light Scattering from Persistent Walks

The transport of vesicles and organelles provide the simplest example of dynamic light scattering from persistent walks. Vesicles and most organelles are much smaller than a wavelength of light and hence represent point scattering objects in motion. In addition, the transport of vesicles and organelles is driven actively. In this section, we describe two ideal models of organelle transport. The first model is constrained and consists of organelles moving on one-dimensional filaments or microtubules. The orientation of these one-dimensional tracks is distributed uniformly in three dimensions. The second model assumes a persistent walk in 3D that is unconstrained. An interesting result of these two models is their non-equivalence: isotropic 3D walks produce different Doppler fluctuation spectra than isotropically distributed 1D walks. These two models can be evaluated in both the extreme limit of very short persistence time (diffusion limit) and the limit of very long persistence time (ballistic limit). The intermediate regime can be approximated by a distribution of lifetime-broadened Doppler spectra to be discussed in the following section.

In dynamic light scattering, coherent speckle is a superposition of the individual partial waves from the individual scattering sources that are in motion. The statistical fluctuations in the speckle intensity are captured by a field autocorrelation function that is obtained as a stochastic sum evaluated using an integral over a probability distribution [54]

$$\begin{aligned}
A_E(\tau) &= \langle E^*(0) E(\tau) \rangle \\
&= E_0^2 + NI_s \int_{-\infty}^{\infty} P(\Delta x) \exp(-i\mathbf{q} \cdot \mathbf{\Delta x}) d\Delta x \\
&= I_0 + NI_s \exp(-iq\langle x \rangle) \left( \frac{1}{2} \int \exp(-q^2 \langle \Delta x^2 \rangle \cos^2 \theta) \sin \theta d\theta \right) \\
&= I_0 + NI_s \frac{\sqrt{\pi}}{2q\Delta x_{rms}} \exp(-iq\langle x \rangle) \, erf(q\Delta x_{rms})
\end{aligned} \quad (15)$$

where $E_0$ is the reference field magnitude, and the field autocorrelation is proportional to the Fourier transform of the probability functional [23]. Eq. (15) is the field-based autocorrelation that would be equivalent to phase-sensitive detection in a dynamic light scattering experiment. There is also an intensity-based autocorrelation function given by

$$\begin{aligned}
A_I(t) &= \langle I^*(0) I(t) \rangle \\
&= N^2 I_s^2 + N^2 I_s^2 \left| \int P(\Delta x) \exp(-i\mathbf{q} \cdot \mathbf{\Delta x}) d\Delta x \right|^2 \\
&= N^2 I_s^2 + N^2 I_s^2 \frac{\pi}{4q^2 \Delta x_{rms}^2} \left| \exp(-iq\langle x \rangle) \, erf(q\Delta x_{rms}) \right|^2
\end{aligned} \quad (16)$$

Field-based autocorrelation is linear in multiple underlying dynamical processes that contribute to the field fluctuations, making interpretations of underlying processes simpler compared with intensity-based autocorrelation. However, the most stable form of fluctuation spectroscopy performed experimentally is with intensity-based detection, because it is less sensitive to mechanical disturbance than the field-based detection (phase-sensitive detection). In the discussion below, field-based descriptions will be used when treating multiple dynamical processes. Intensity-based descriptions will be used for experimental studies and for pure theoretical cases with simple limiting behavior when persistent walks along isotropically-oriented filaments or microtubules are driven by molecular motors that run at approximately constant speeds but with a distribution of persistence times. There are three limiting cases: (1) diffusive motion in 1D, (2) diffusive motion in 3D, and (3) ballistic motion. In all three cases, the motion is averaged isotropically over all angles.

1D Isotropic Diffusion Limit

One-dimensional isotropic transport is a model for which particles are confined in one direction, with both positive and negative excursions along a line, while the direction is distributed isotropically in 3D. The distribution function for one-dimensional isotropic motion is [27]

$$P(\Delta x) = \frac{1}{\sqrt{4\pi Dt}} \exp(-\Delta x^2 / 4Dt) \tag{17}$$

The intensity autocorrelation function can be written as

$$A_I(t) = N^2 I_s^2 + N^2 I_s^2 \left( \frac{1}{4\pi} \iint P(\Delta x) \exp(-iq\Delta x \cos\theta) \mathrm{d}\Delta x \mathrm{d}\Omega \right)^2$$

$$= N^2 I_s^2 + N^2 I_s^2 \left( \frac{1}{2} \int \exp(-q^2 Dt \cos^2\theta) \sin\theta \mathrm{d}\theta \right)^2 \tag{18}$$

$$= N^2 I_s^2 + N^2 I_s^2 \frac{\pi}{4q^2 Dt} \mathrm{erf}^2\left(\sqrt{q^2 Dt}\right)$$

The autocorrelation function behaves as the error function with the characteristic time $1/q^2 D$.

3D Isotropic Diffusion Limit

Three-dimensional isotropic transport is the model for which particles are free in 3 dimensions. The distribution function of the three-dimensional isotropic motion is

$$P(\Delta r) = \left(\frac{1}{\sqrt{4\pi Dt}}\right)^3 \exp(-\Delta x^2 / 4Dt) \exp(-\Delta y^2 / 4Dt) \exp(-\Delta z^2 / 4Dt) \tag{19}$$

The intensity autocorrelation function is

$$A_I(t) = N^2 I_s^2 + N^2 I_s^2 \left( \iiint_{\Delta r} P(\Delta x) \mathrm{d}\Delta x P(\Delta y) \mathrm{d}\Delta y P(\Delta z) \exp(-iq\Delta z) \mathrm{d}\Delta z \right)^2$$

$$= N^2 I_s^2 + N^2 I_s^2 \exp(-2q^2 Dt) \tag{20}$$

The autocorrelation function is an exponential equation, and the characteristic time is $1/q^2 D$ that is the same as for the one-dimensional isotropic model. The spectral density, calculated using the Wiener-Khinchin theorem, is

$$S(\omega) = FT[A_I(t)](\omega) = \sqrt{2\pi} N^2 I_s^2 \left[ \delta(\omega) + \frac{1}{\pi} \frac{2q^2 D}{(2q^2 D)^2 + \omega^2} \right] \quad (21)$$

The second term in the spectral density is a typical Lorentzian function. In the low and high frequency limits, these are

$$S(\omega) = \begin{cases} \dfrac{\sqrt{\dfrac{2}{\pi}} N^2 I_s^2}{2q^2 D} = \text{const}, & \omega \ll 2q^2 D \\[2ex] \dfrac{2\sqrt{\dfrac{2}{\pi}} N^2 I_s^2 q^2 D}{\omega^2} \propto \omega^{-2}, & \omega \gg 2q^2 D \end{cases} \quad (22)$$

1D and 3D Isotropic Ballistic Limit

In the ballistic limit of long persistence time, the three-dimensional ballistic transport model is identical to isotropically distributed 1D transport, so they share the same limit. The displacement is $\Delta r = vt$, and the distribution function for this type of motion is

$$P(\Delta r) = \delta(\Delta r - vt) \quad (23)$$

The intensity autocorrelation function is

$$\begin{aligned} A_I(t) &= N^2 I_s^2 + N^2 I_s^2 \sum_{i \neq j} \sum_j \exp(-iqvt\cos\theta_i)\exp(iqvt\cos\theta_j) \\ &= N^2 I_s^2 + N^2 I_s^2 \left( \frac{1}{4\pi} \int \exp(-iqvt\cos\theta) d\Omega \right)^2 \\ &= N^2 I_s^2 + N^2 I_s^2 \text{sinc}^2(qvt) \end{aligned} \quad (24)$$

where the oscillatory sinc function arises from the ballistic Doppler frequency. The spectral density is

$$S(\omega) = FT[A_I(t)](\omega) = \sqrt{2\pi} N^2 I_s^2 \left[ \delta(\omega) + \frac{1}{2qv} \text{tri}\left(\frac{\omega}{2qv}\right) \right] \quad (25)$$

where $\text{tri}(x)$ is the triangular function.

Information contained within the autocorrelation function is contained equivalently within the spectral power density, but when there are many subensembles contributing to the dynamic light scattering, and the characteristic time scales are widely separated across several orders of magnitude, the fluctuation spectrum is a more "natural" representation than the autocorrelation by separating out processes according to their respective characteristic frequencies. For instance, when the subcellular transport is quasi-ballistic, the fluctuation frequencies of the fluctuation spectra are closely related to the Doppler frequencies of the moving scatterers. Doppler fluctuation power spectra, even in the homogeneous case, have no spectral peak, but are fluctuation spectra with zero mean frequency and characteristic "edge" or "knee" frequencies, as shown in Fig. 5(a) for the three limiting cases: ballistic, 3D diffusion, and isotropic 1D diffusion. On a logarithmic frequency scale, the diffusive fluctuation spectra show a characteristic "roll-off" of a Lorentzian lineshape of zero mean frequency, while the ballistic spectrum displays a "Doppler edge" above which the fluctuation spectral power density drops rapidly.

Intermediate Cross-Over Regime

Between the diffusive and the ballistic limits is the cross-over regime when $N_D \approx 1$ , with the mean free path $L_p$ in the range of 50 nm for a wavelength at 840 nm in the infrared using a backscattering optical configuration. The cross-over regime, with significant deviations from the ideal limits, is relatively wide, with the mean free path spanning from approximately 10 nanometers to a quarter of a micrometer.

An ensemble of $N$ particles executing persistent walks with an exponential distribution of persistence time $t_p$, inclined at angle $\phi$, and with no discontinuous phase jumps between walk segments, produce a characteristic damped-harmonic oscillator power spectrum with a lineshape given by

$$S_E(\omega,\phi) = 2N\pi^2 \frac{\omega_\phi^2 \gamma}{\left(\omega^2 - \omega_\phi^2\right)^2 + \omega^2 \gamma^2} \tag{26}$$

where the damping factor is inversely related to the mean persistence time $t_p$ through $gt_p = 1$. In an isotropic tissue, the colinearity angles are distributed as

$$P(\phi)\mathrm{d}\phi = \sin\phi \mathrm{d}\phi \tag{27}$$

For a distribution of Doppler frequencies $\omega_\phi$ caused by the distribution of angles, the increment to the fluctuation power spectrum is

$$\begin{aligned}\mathrm{d}S_E(\omega) &= L(\omega,\phi)P(\phi)\mathrm{d}\phi \\ &= \frac{\gamma N}{\pi}\left[\frac{\omega_D^2 \cos^2\phi}{\left(\omega_D^2 \cos^2\phi - \omega^2\right)^2 + \omega^2 \gamma^2}\right]\sin\phi \mathrm{d}\phi\end{aligned} \tag{28}$$

For the frequency distribution from an isotropic tissue, the total power spectrum is integrated over all Doppler frequencies as

$$S_E(\omega) = \frac{\gamma \omega_D^2 N}{\pi} \int_0^\pi \left[ \frac{\cos^2\phi}{\left(\omega_D^2 \cos^2\phi - \omega^2\right)^2 + \omega^2\gamma^2} \right] \sin\phi \, d\phi$$

$$= \frac{\gamma N}{\pi \omega_D} \int_{-\omega_D}^{\omega_D} \left[ \frac{y^2}{\left(y^2 - \omega^2\right)^2 + \omega^2\gamma^2} \right] dy \quad (29)$$

Examples of isotropically-averaged 1D motion are shown in Fig. 5(b) for Doppler numbers $N_D$ = 0.1, 1, and 10. The dashed curves are for unidirectional 1D motion, showing a clear Doppler peak at $f_D = 1$ Hz for $\omega_D t_p = 10$. In the $\omega_D t_p = 0.1$ case, there is a diffusion knee at $f_d = q^2 v_0^2 \tau / 2\pi \approx 0.1$ Hz. The cross-over regime is captured when $\omega_D t_p = 1$. The isotropic averaging produces a fluctuation spectrum that has no peak at the Doppler frequency, even in the case of large $N_D$, although there is a distinct edge at the Doppler frequency for this case. When the Doppler number is small, a diffusive knee structure emerges at lower frequencies.

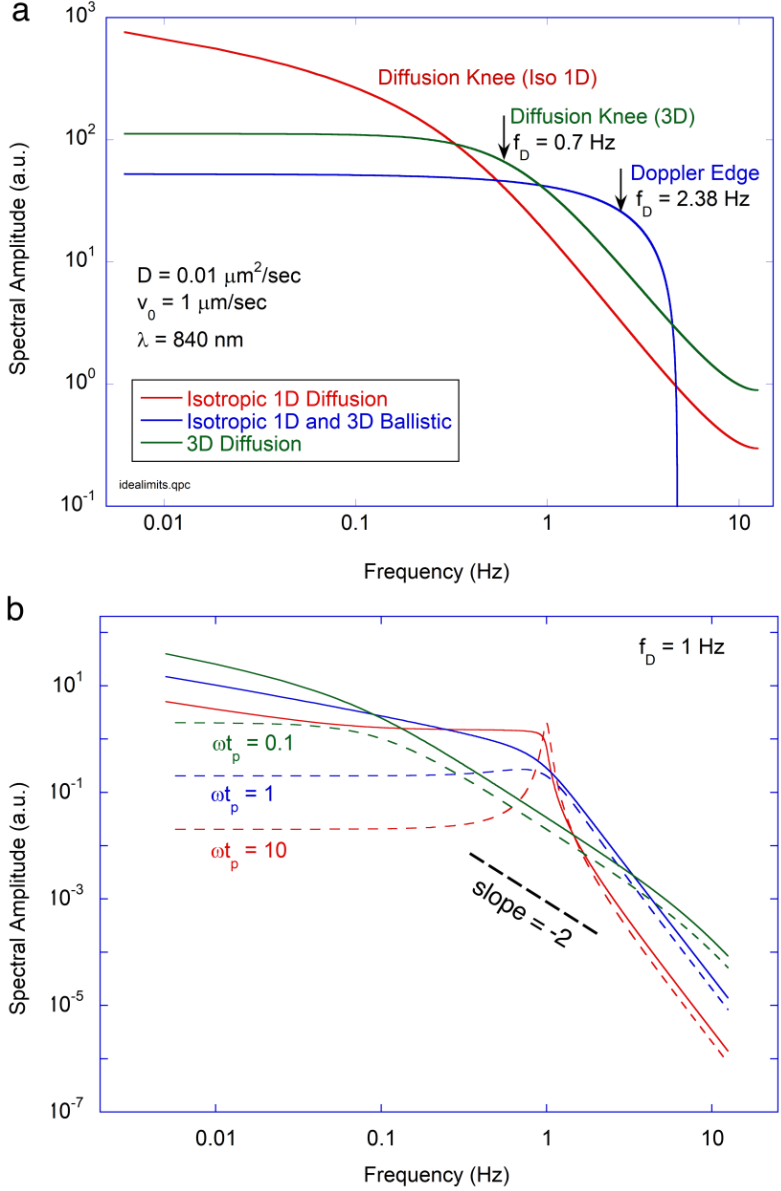

Fig. 5 (a) Fluctuation spectra for three limiting cases: Isotropic 1D diffusion, 3D diffusion and the ballistic case averaged isotropically over all angles. A diffusion coefficient is used for the first two cases, and a uniform velocity is applied to the last case. (b) Comparison of unidirectional (dashed) versus isotropically averaged 1D (solid) power spectra when the $N_D = 0.1$, 1, and 10.

The cross-over behavior from the Doppler regime to the diffusion regime is described in terms of a knee frequency, which is a function of diffusion and ballistic frequencies and persistence time

$$\omega_{\text{knee}} = \frac{\omega_D^2}{\sqrt{1/t_p^2 + \omega_D^2}} = \frac{\omega_{\text{diffusion}}}{\sqrt{1 + \omega_D^2 t_p^2}} \tag{30}$$

For long persistence times $t_p$, the knee frequency is a Doppler edge that is associated with a slope < -2, while for short persistence times, the knee frequency is the diffusive roll-off frequency $\omega_{\text{diffusion}} = q^2 D$. The knee frequency is shown in Fig. 6 as a function of the mean intracellular speed for a range of persistence times, assuming no correlation between mean speed and mean persistence time. However, most biological processes display correlations between speeds and persistence times. The simplest scaling for such a correlation is $vt_p = L_p$, as discussed in Section 2.2.

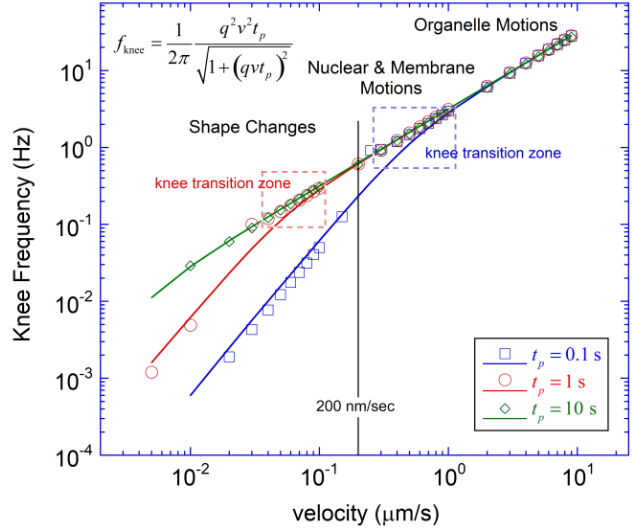

Fig. 6 Knee frequency versus the mean intracellular speed for persistence times ranging from 0.1 seconds to 10 seconds. The markers are knee frequencies extracted from zero points of $d^3(\log S)/d(\log f)^3$ which are related to change in curvature, and lines are plots of Eqn. (30). The values of S are numerical calculations from Eqn. (29). The region labeled "knee transition zone" is when more than one knee appears in the 0.01 to 12.5 Hz range. The knee frequencies of living tissue range from 0.01 Hz to 1 Hz, corresponding to speeds from 3 nm/sec (cellular shape changes) to 300 nm/sec (nuclear and membrane motions).

### *3.2 Monte Carlo Simulation of Transport and Light Scattering*

The theoretical predictions were compared to Monte Carlo DLS simulations to validate the theoretical model for isotropically-averaged one-dimensional transport processes. A calibration simulation was performed first on transport in the diffusion limit to test the three-dimensional diffusion case in contrast to the isotropically-averaged one-dimensional diffusion case. The Monte Carlo simulations were performed with 5000 particle walkers that contribute coherent scattered waves to the far field where the net complex field is sampled at a chosen sampling rate and transformed to the frequency domain through a fast Fourier transform. The resulting complex-valued fluctuation spectrum is taken modulus-squared and averaged over an ensemble of 50 simulations. The fluctuation spectral power density, in this case, is in the "heterodyne" mode to be compared with theoretical calculations. The walkers were simulated with a diffusion coefficient of $0.002\ \mu m^2/s$ using a probe wavelength of $0.84\ \mu m$. The sampling frame rate was 25 frame/s, and the capture is assumed to be instantaneous (infinitesimal exposure time). The total capture time is either 100 sec or 200 sec.

Fig. 7(a) shows the theoretical calculation and simulation of the Wiener process in one-dimensional isotropic and three-dimensional isotropic transport. Numerical calculations are derived from the autocorrelation function via the Wiener-Khinchin theorem from Eq. (18) and Eq.(21), respectively. Fig. 7(b) shows three cases: the two limits with persistence times $t_p \to 0$, $t_p \to \infty$, and for a moderate persistence time $t_p = 0.5$ s in the cross-over regime, with persistence times distributed according to an exponential probability.

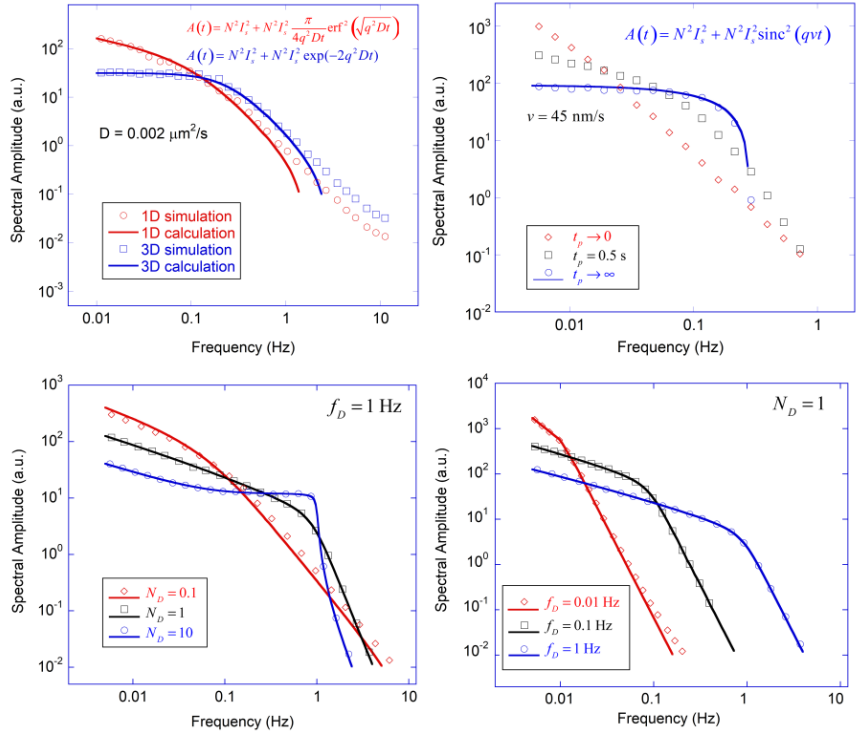

Fig. 7 (a) Theoretical calculations (solid curves) and Monte Carlo simulations (markers) of three-dimensional isotropic transport versus one-dimensional isotropically-averaged transport. The theoretical calculation is from the Fourier transform of the autocorrelation function. The three-dimensional isotropic transport has a higher knee (0.8 Hz) than the one-dimensional isotropic transport (0.05 Hz). The autocorrelation functions are an exponential and an error function, respectively. In the 3D case, the low-frequency limit is flat and the high frequency has a -2 slope, which agrees with limits in Eq. (22). The high-frequency discrepancy in the calculated curve is a numerical artifact originating from the finite sampling of the autocorrelation function. (b) Monte Carlo simulation of 1D isotropic persistent walk in three regimes and theoretical results in the ballistic regime from the autocorrelation function. (c) – (d) Monte Carlo simulation and numerical calculation of persistent walks in the intermediate regime: (c) with fixed Doppler frequency $f_D$. (d) with fixed Doppler number $N_D$=1.

Similar Monte Carlo simulations in the cross-over regime were carried out that match closely with the analytical result from Eq. (29) [Fig. 7(c)-(d)], although the high-frequency side of the spectrum has a Nyquist floor. The simulations were done for conditions similar to experimental measurements (discussed in the next section), and the finite-time sampling means that walk events that are long compared to the persistence time may end outside the observation timeframe, being recorded as an event with a shorter time. As a result, the spectrum starts to flatten above the frequency around $1/t_p = 2\pi f_D / N_D$, and the effect is more visible in processes with a longer mean persistence time (lower $f_D$ or larger $N_D$), because of the exponential distribution of persistence times.

*3.3 Doppler Fluctuation Spectra of Living Tissues*

Fig. 8 shows examples of tumor spheroid spectra and their spectral responses to drugs measured with the BDI system. Most spectra have a Lorentzian-like shape, with a knee at low frequency, a power-law roll-off in the mid frequency and a floor near the Nyquist frequency. Fig. 8 (a) compares the spectrum of a PaCa2-derived spheroid to an HT29-derived spheroid, showing a higher Doppler knee frequency in the case of the more loosely aggregated PaCa2 spheroid. In Fig. 8 (b) paclitaxel is applied to a DLD-1 spheroid. The cytoskeletal drug stabilizes polymerization of tubulin, lowers the rates of microtubule dynamic instability in human tumor cells [55] and causes cell death [56, 57]. The Doppler knee shifts to lower frequency caused by the increased stiffness of the cell. This represents a "red shift" in frequency content. In Fig. 8 (c) valinomycin, a mitochondrial ionophore, facilitates $K^+$ charge movement and triggers loss of mitochondrial membrane potential, DNA fragmentation and death [58, 59]. The spectrogram pattern observed in this case is correlated with apoptosis [60]. In Fig. 8 (d) the relative change in spectral content for valinomycin is displayed as a relative spectrogram with frequency along the horizontal axis and time along the vertical axis. The spectral change is relative to the average baseline spectrum (average of spectra prior to the application of the drug at $t = 0$). The spectrogram displays a suppression of the Doppler edge while enhancing high-frequency and low-frequency content.

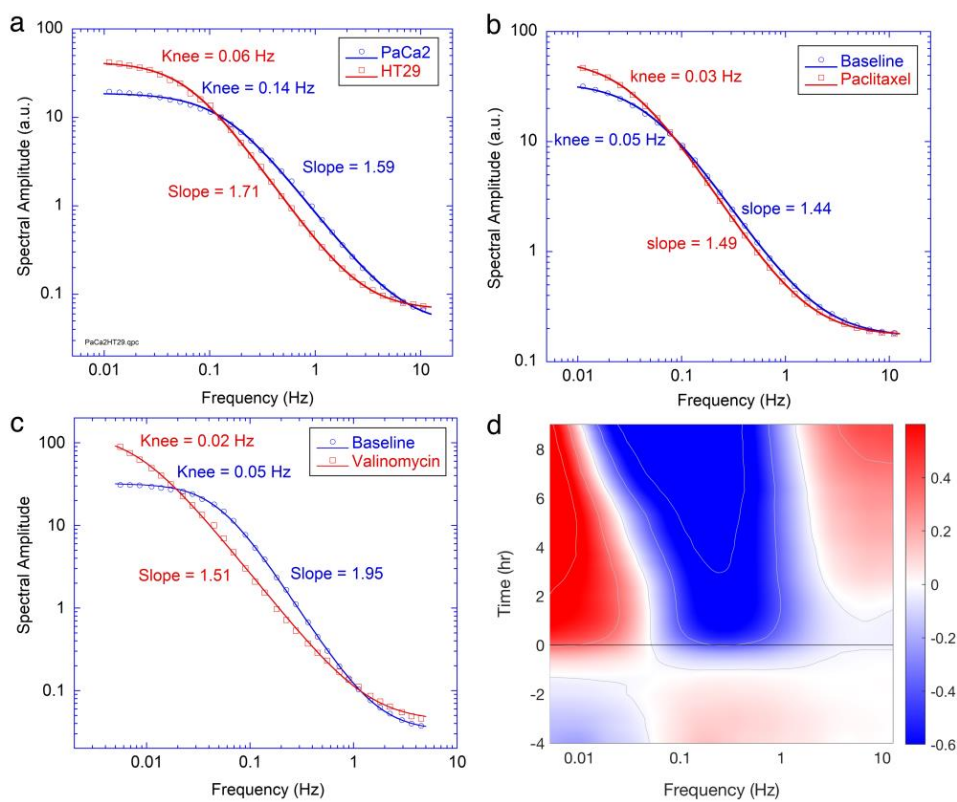

Fig. 8 Examples of fluctuation power spectra. Markers are experimental data and solid curves are guiding lines. Knee and slope values are approximate numbers from curve fitting. (a) PaCa2-derived spheroids form a loose aggregate of cells and display a higher Doppler knee frequency than HT29-derived spheroids that form compact spheroids with tight and dense cellular adhesions. (b) DLD-1 spheroid responding to 50 μM paclitaxel. (c) The effect of valinomycin, a mitochondrial drug, on a DLD-1 spheroid. The baseline (pre-drug) spectrum shows a strong Doppler knee that is suppressed under the application of 50 μM valinomycin. (d) The relative change in spectral content in a spectrogram (time-frequency) format for the case of valinomycin. The drug is applied at $t = 0$, suppressing the Doppler knee.

Compared with cell-line spheroids, tumor biopsies show more heterogeneity among samples and more diverse responses to treatment. The biopsy samples obtained from resected tissue or needle cores were carefully dissected by hand to avoid connective tissues or fat, which have relatively low activity. Biopsies display spatial heterogeneity in the dynamics, including motility, spectrum and spectral responses. In a study on a standard-of-care chemotherapy treatment (cyclophosphamide, doxorubicin, prednisolone and vincristine) of dogs with B cell lymphoma, lymph node biopsies were treated with the combination treatment as well as by the single-agent compounds. The averaged spectral response of canine biopsy tissue resistant to vincristine is shown in Fig. 9 (a) and (b) [53]. Vincristine is a *vinca* alkaloid that prevents polymerization of tubulin and induces depolymerization of microtubules, blocking mitosis during metaphase by arresting cells, and causing cell death by apoptosis. The spectrogram displays an enhanced mid-frequency in response to the drug which may be a marker for drug resistance. The spectra in Fig. 9 (c) are biopsies from two different esophageal cancer patients. Patient-1 has low activity with no discernible Doppler edge. However, the biopsy from Patient-2 displays a distinct Doppler edge near 0.2 Hz that becomes sharper after the addition of carboplatin, a DNA drug that leads to apoptosis. The power spectrum has an almost flat power density at low frequencies, with a distinct Doppler edge and a large negative slope of $s = -2.4$.

The associated spectrogram for Patient-2 is shown in Fig. 9 (d). The sharpening of the Doppler edge appears as a dark red strip in the mid-frequency range. These data are consistent with the existence of a Doppler edge in these patient samples. As pointed out in Eq. (14), an experimental spectrum is an envelope of Doppler broadened spectra of processes with different $N_D$ and $f_D$ values. As a result, sharp Doppler edges or knees of individual processes are washed out, and there is not a well-defined single $N_D$ or $f_D$ value for an experimental spectrum. However, a spectrum with a Doppler edge and a steep slope is strongly indicative of processes with high $N_D$ and $f_D$ values and highly ballistic motions.

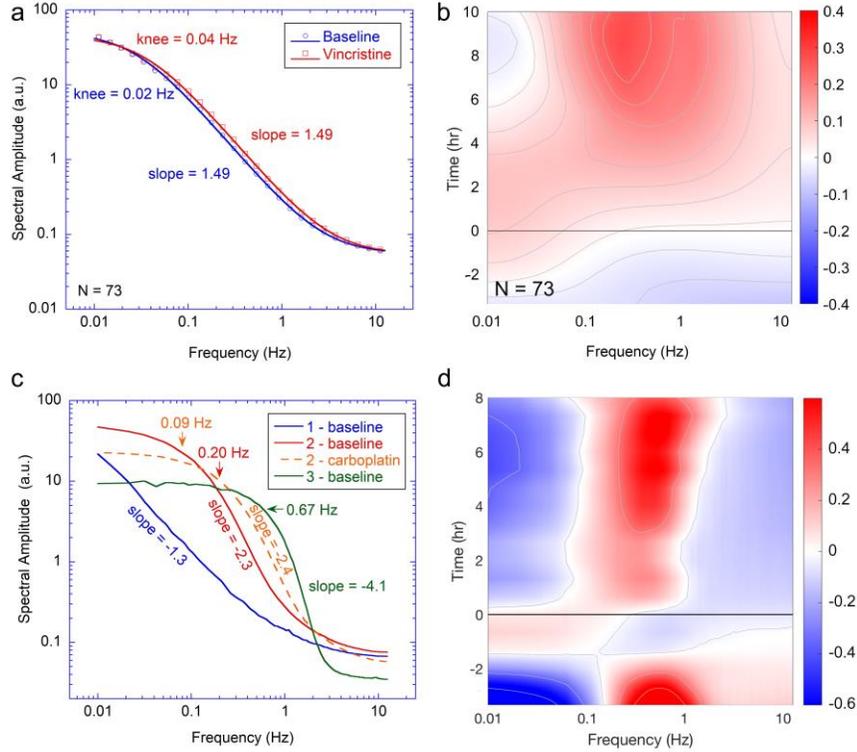

Fig. 9 Examples of spectral responses for living biopsy samples treated with anti-cancer drugs. (a) Canine B-cell lymphoma biopsies responding to 60 nM vincristine from dogs that are resistant to chemotherapy. The final spectrum is 10 hours after application of the drug compared against baseline. (b) A time-frequency spectrogram response associated with (a) shows enhancement in mid frequency. The spectrogram displays the net effect compared against control medium DMSO. (c) Effect of 25 μM carboplatin drug on two *ex vivo* biopsy samples for esophageal cancer from different patients, plus the baseline spectrum of a third sample. The spectral shape for sample-1 is almost linear at low-mid frequency, while sample-2 and -3 have a spectrum with a sharp Doppler edge. (d) Time-frequency spectrogram of sample-2 from (c) showing the emergence of a sharper Doppler edge.

## 4. Discussion

Fluctuation power spectra from living tissue display characteristic spectral shapes that are reminiscent of the common diffusive power spectra obtained from dynamic light scattering measurements of diffusing particles. This has led to conventional interpretations that consider intracellular transport to be primarily in the diffusive regime without strong ballistic character. However, this tentative conclusion from dynamic light scattering contradicts a vast literature from two-dimensional cell culture that directly tracks motions with long persistence lengths that places most active intracellular processes in the ballistic regime. The resolution of this contradiction is simply the superposition of many ballistic processes in living tissue with a wide range of characteristic frequencies.

We have investigated the fluctuation spectra of transport modeled by persistent walks in the dynamic light scattering setting. The Doppler number defined as $N_D = qvt_p$ is a dimensionless scaling parameter that determines the regime of the motion and the spectrum shape. Many intracellular motions, including the processes associated with kinesin, dynein, and filopodia, have a long persistence length, leading to Doppler numbers greater than 1, placing the motions in the Doppler regime. In the intermediate regime, the power spectrum of a damped harmonic oscillator averaged over all angles yields the Doppler spectrum while in the ballistic and diffusive regimes, the power spectra are obtained through Fourier transforms of autocorrelation functions of intensities.

Our model builds a framework for interpreting fluctuation spectra. A sample spectrum can be understood as a sum of processes with different Doppler frequency shifts (or velocities) and Doppler numbers (or persistence lengths). The slopes of spectra at high frequency for the ballistic and diffusion limits are -∞ and -2, respectively, indicating that a slope steeper than -2 is characteristic of persistent walks. In spheroid and biopsy spectra, the greater the (absolute) slope, the further the motions deviate from diffusive behavior, with walks having longer persistence lengths. It is interesting to note that many metabolically-active tumor spheroids and biopsies show a typical spectral slope parameter of $s = -1.7$. If the typical Doppler number for active processes is assumed to be $N_D > 3$ (with a sharp Doppler edge), then the probability density function needed to yield a slope parameter of -1.7 would have $(1/f)^k$ character with approximately $k \approx 0.6$. Therefore, the spectral contributions to the fluctuation spectra increase at lower frequencies, consistent with stronger light scattering from membranes and cell-scale optical heterogeneities.

Experimental evidence for the Doppler edge is obtained using biodynamic imaging (BDI). BDI is a coherent imaging technique that records the field information from backscattering and generates Doppler fluctuation spectra. For a given sample, the spectrum change caused by the addition of an anti-cancer drug can be understood as the speed up or slow down of certain processes. A 10-hour time-lapse measurement of drug response captures the change of the velocities over time. These shifts may eventually be correlated to specific drug mechanism, providing insights for treatment and drug development.

Active transport processes in cells often are described by variations on the random walk. For instance, a Lévy flight is a random walk where the lengths of individual jumps are distributed with a probability density function $P(x) \propto |x|^{-\alpha-1}$ when $x$ is large [61]. Levy and Cauchy flights produce anomalous diffusion because they have "fat tail" distributions with no finite variance [62]. Conversely, in the continuous-time random walk (CTRW) model [63] a particle waits between jumps for times set by a distribution function that also may have fat tails, producing anomalous subdiffusion. Combining waiting-time with jump-length models produces anomalous diffusion tunable continuously from subdiffusive to superdiffusive behavior. Future work will investigate this anomalous regime.

While the samples used in this paper are tumor biopsies and spheroids, the light scattering analysis can be extended to other forms of life. Swimming bacteria have transport known as "run and tumble". Given that the velocity of a bacterium is 2 μm/sec to 200 μm/sec [64], the motion is firmly in the Doppler regime, and a sharp edge is expected in the spectrum, which would be suppressed if the bacteria slow down. In addition, cell divisions in gametocytes and zygotes may be slow processes that take place over a few hours, but they are firmly in the Doppler regime. Furthermore, fluctuation spectra of biased random walks, Levy walks, Cauchy walks, etc. can be studied, producing characteristic shapes and features that can help understand experimental observations. Therefore, biodynamic imaging and intracellular Doppler spectroscopy are poised to provide new insight into tissue dynamics and potentially important new screens of drug mechanisms.

**Funding**

National Science Foundation (NSF) (1263753-CBET); and the National Institutes of Health (NIH) (R01-EB016582) and (R01-HD078682).